\documentclass[%showpacs, 
superscriptaddress,twocolumn, preprintnumbers, nofootinbib,amsmath,amssymb]{revtex4}

\usepackage{amssymb}
\usepackage{amsmath}
\usepackage{graphicx}
\usepackage{longtable}
\usepackage{verbatim}
\usepackage{amsfonts}

\newcommand{\be}{\begin{equation}}
\newcommand{\ee}{\end{equation}}
\newcommand{\e}{\epsilon_*}
\newcommand{\bea}{\begin{eqnarray}}
\newcommand{\eea}{\end{eqnarray}}
\newcommand{\beq}{\begin{equation}}
\newcommand{\eeq}{\end{equation}}
\def\beqa{\begin{eqnarray}}

\def\d{{\rm d}}
\def\eeqa{\end{eqnarray}}

\def\lsim{\mathrel{\rlap{\lower4pt\hbox{\hskip0.5pt$\sim$}}
    \raise1pt\hbox{$<$}}}         %less than or approx. symbol
\def\gsim{\mathrel{\rlap{\lower4pt\hbox{\hskip0.5pt$\sim$}}
    \raise1pt\hbox{$>$}}}         %greater than or approx. symbol

\def\mp{M_{\rm p}}
\begin{document}

\vspace*{-30mm}

\title{What We Can Learn  from  the Running of the Spectral Index 
\vskip 0.1cm
if no Tensors are Detected in the Cosmic Microwave Background Anisotropy
}

\vskip 0.3cm

\author{M. Biagetti}
\affiliation{D\'epartement de Physique Th\'eorique and Centre for
  Astroparticle Physics (CAP), Universit\'e de Gen\`eve, 24 quai E. Ansermet, CH-1211 Geneva, Switzerland
}
\author{A. Kehagias}
\affiliation{D\'epartement de Physique Th\'eorique and Centre for
  Astroparticle Physics (CAP), Universit\'e de Gen\`eve, 24 quai E. Ansermet, CH-1211 Geneva, Switzerland
}
\affiliation{ Physics Division, National Technical University of Athens, 15780 Zografou Campus, Athens, Greece}

\author{A. Riotto}
\affiliation{D\'epartement de Physique Th\'eorique and Centre for
  Astroparticle Physics (CAP), Universit\'e de Gen\`eve, 24 quai E. Ansermet, CH-1211 Geneva, Switzerland
}
%\affiliation{ Physics Division, National Technical University of Athens, 15780 Zografou Campus, Athens, Greece}

\date{\today}

%\title{ Title\\
%\vspace*{0.2cm}
%Title}
%
%\vspace*{2cm}
%\author{ Alex Kehagias }
%\author{ Antonio Riotto }
%
%%\email{antonio.riotto@unige.ch}
%\affiliation{D\'epartement de Physique Th\'eorique and Centre for
%  Astroparticle Physics (CAP),\\
%  \vspace*{0.1cm}
%Universit\'e de Gen\`eve, 24 quai E. Ansermet, CH-1211 Gen\`eve, Suisse}
%\vspace*{1cm}

\begin{abstract}
\noindent
In this paper we  operate under the  assumption that no tensors from inflation  will be measured in the future by the dedicated experiments and argue that, while for single-field slow-roll models of inflation the running of the spectral index will be hard to be detected, in multi-field models the running can be large due to its  strong correlation with   non-Gaussianity. A detection of the  running 
might therefore be related to the presence of   more than one active scalar degree of freedom during inflation.
\end{abstract}

\maketitle
\noindent
{\em Motivations.}--- 
%%%%%%%%%%%%%%%%%%%%
Inflation \cite{guth81} provides a mechanism 
to explain the 
initial conditions for the Large Scale Structure (LSS) and for the Cosmic
Microwave Background (CMB) anisotropy. The seeds for the  density  scalar perturbations are
created  from the quantum fluctuations ``redshifted'' out of the Hubble radius  during the 
early period of superluminal expansion of the universe.  
At the same time, the  generation of tensor (gravity-wave) fluctuations 
is a generic prediction of an accelerated  de Sitter expansion of the universe \cite{lrreview}.

To characterize the scalar and tensor perturbations several observables are introduced: the spectral index (or tilt) $n_s$ of the scalar 
perturbations, currently measured  to be $n_s=0.9655\pm 0.0062$ at 68\% CL; the running with the wavenumber $k$ of the spectral index $\alpha_s=\d n_s/\d\ln k$, currently consistent with zero,  $\alpha_s=-0.00571\pm 0.0071$ at 68\% CL (assuming no tensors); and the level of local non-Gaussianities parametrized by $f_{\rm NL}=0.8\pm 5.0$ at 68\% CL  and $g_{\rm NL}=(-9.0\pm 7.7 )\times 10^4$ at 68\% CL. 
All these figures are quoted from the last Planck releases \cite{Planck15a,Planck15b}. Finally, the tensor-to-scalar ratio is bounded to be $r<0.12$ at  95\% CL \cite{combined}.

Many next-generation satellite missions, see for instance Refs.  \cite{core,EPIC2m,LiteBIRD}, are dedicated to   measure the polarization of the CMB  anisotropies. On large angular scales the B-mode polarization of the CMB carries the imprint of primordial gravitational waves, and its precise measurement would provide a powerful probe of the epoch of inflation. The goal of these missions  is to achieve a measurement of  $r$ down to ${\cal O}(10^{-3})$. 

While a detection of gravity waves from inflation will be a major milestone for the inflationary paradigm, on the other hand one needs to remember that going from $r=10^{-1}$ down to $r=10^{-3}$ means
testing  the energy scale of inflation $E_{\rm inf}$ only by a factor $(100)^{1/4}\simeq 3.1$ better than current bounds 
($r$ scales like $E_{\rm inf}^4$). 

One should therefore envisage the situation in which Nature has not be so kind to us to set the scale of the energy of inflation within a factor of three away from the current limits. In this paper we will therefore operate under the working assumption that no tensors will be measured in the future by the dedicated experiments. 

While we hope  this hypothesis will  prove to be wrong in the future, we believe it is reasonable to ask how   else we could learn something  more about the inflationary perturbations.   

In this paper we turn our attention to  the scale dependence of the tilt of the scalar perturbations.
Our hypothesis on the tensor modes leads to a hardly detectable running within single-field slow-roll models. On the other hand,  a negative result on  tensors might indicate a low level of the Hubble rate, leaving  open the possibility that the  scalar perturbations are due to a light scalar field other than the inflaton field. In such a case, our findings indicate that 
a large running may be achieved thanks to an interesting  correlation between  the running 
and the level of non-Gaussianity. 
Such strong correlation exists only in multi-field models of inflation, while it is missing in single-field slow-roll models. If no tensors will be observed, a measurement of a sizable running of the spectral index
might therefore imply a sizable amount of non-Gaussianity (and viceversa) and the presence of more than one active scalar degree of freedom during inflation.

Our considerations are particularly relevant  for the goals of next  missions  designed to measure  the three dimensional
structure of the universe, such as 
the ESA mission EUCLID \cite{euclid} and 
the NASA SPHEREx mission \cite{dore}, a proposed all-sky spectroscopic survey satellite. For instance,  the  combination of the galaxy power spectrum and bispectrum leads to the forecasted 68\% CL errors   $\sigma(n_s-1)=2.2\times 10^{-3}$, $\sigma(\alpha_s)=6.5\times 10^{-4}$  and $\sigma(f_{\rm NL})=0.2$ (for the local case) in the case of SPHEREx.  This is  an improvement with respect to EUCLID forecasts of a factor $\sim 2$ in $\alpha_s$  and a factor 
%local $f_{\rm NL}$
$\sim 20$  for the  local $f_{\rm NL}$ (or $\sim 6$ with the galaxy power spectrum only).
For the other parameters SPHEREx will deliver constraints comparable to those expected from
the EUCLID spectroscopic survey. 
%better than the expected constraint from the EUCLID \cite{euclid}. 

As EUCLID and (possibly) SPHEREx will strongly narrow down the allowed parameter space in the $(\alpha_s,f_{\rm NL})$-plane, it is
therefore  extremely interesting and timely to
understand what are the theoretical expectations in  terms of $\alpha_s$ and its connections to  non-Gaussianity.

\vskip 0.2cm
\noindent
{\em Single-field slow-roll inflation.}--- Let us first  consider the case in which the cosmological perturbations are generated by the same scalar field $\phi$ driving
inflation. The power spectrum of scalar perturbations is given in such a case by  

\begin{equation}
{\cal P}_s(k)=
\frac{1}{2 \mp^2\e}\left(\frac{H_*}{2\pi}\right)^2,
%\left(\frac{k}{a_*H_*}\right)^{n_s-1},
\label{pscalar}
\end{equation}
where the sub-index ${}_*$ indicates that quantities have to be computed at Hubble crossing, $H_*$ is the Hubble rate during inflation,
$\mp\simeq 2.4\times 10^{18}$ GeV is  the reduced Planck mass, and $\e=-\dot{H}_*/H_*^2$ is one of the slow-roll parameters (dots  indicate derivatives with respect to time).

%The spectrum of tensor modes is given by
%
%\begin{equation}
%{\cal P}_{T}(k)=\frac{8}{\mp^2}\left(\frac{H_*}{2\pi}\right)^2
%\left(\frac{k}{aH_*}\right)^{n_T},
%\end{equation}
%where $n_T=-2\e$ is the tensor spectral index. 

Let us now assume that no tensors will  be  observed in the future and see what are the consequences we can draw. 
Measuring no tensors   implies that $r=16\e\ll 10^{-3}$ or $\e\ll 10^{-4}$.
Irrespectively of the form of the inflaton potential,
this means that  $\phi\ll \mp$.  To see this, take the cosmological range of scales to span four
decades, corresponding to $\ln k\sim 10$. This corresponds to $\sim 10$ {\it e}-folds of inflation. In first approximation in   slow-roll
inflation $\e$ has negligible variation over one {\it e}-fold and in typical models it has only small
variation over the $\sim 10$ {\it e}-folds. Taking that to be the case,  one finds (this is nothing else that the reverse of the so-called Lyth bound \cite{lythgrav}) $\phi\ll 0.4\,\mp$, 
%
%\be
%\phi\ll 0.4\,\mp,
%\ee
%\be
%\e\ll\frac{1}{2}\left(\frac{1}{9}\right)^2\simeq 6\times 10^{-3},
%\ee
where we have used (primes indicate derivative with respect to the inflaton field) $
 N'(\phi)=\pm(1/\sqrt{2\epsilon}\mp)$ and $N$ is the number of {\it e}-folds till the end of inflation.
  A  more detailed computation leads to sub-Planckian excursions  of the inflation field for $r\lsim 2\times 10^{-5}$ \cite{roest}.
 
 The first consequence of observing no tensors is therefore that
the  model of inflation will be  characterized by  a potential  of the form $V(\phi)=V_0+\cdots$  with the
constant first term dominating. 

The second consequence is that one can simplify the formulae for the spectral index and its running
 \cite{lrreview}

\be
n_s-1=2\eta_*,\,\,\,\,\,\,\,
\alpha_s=2\xi_*^2,
\ee
%\begin{eqnarray}
%n_s-1&=&2\eta_*,\nonumber\\
%\alpha_s&=&2\xi_*^2,
%\end{eqnarray}
where $\eta_*=\mp^2V''/V$ and $\xi_*^2=\mp^4(V' V'''/V^2)$.
Therefore, observing no tensors implies that the running of the spectral index will be sizable only if the third derivative of the potential is nonvanishing. Also, since  $(n_s-1)=2\eta_*$ is measured to be negative by Planck, we conclude that the  second derivative of the potential must be negative.

A prototype of inflationary models summarizing all these properties (domination of   the vacuum energy, a nonvanishing $V'''$ and $V''<0$)   is represented by the following form of the potential ($c>0$)

\be
\label{proto}
V(\phi)=V_0\left(1-c\phi^p\right),\,\,\,\,   -\infty<p<0 \,\,{\rm or}\,\,  p>1.
\ee
This parametrization has the virtue to reproduce for $p\rightarrow -\infty$  the exponential models $V=V_0\left(1-e^{-q\phi}\right)$, among which the popular Starobinsky ${\cal R}^2$-model \cite{star} and the Higgs inflation model \cite{higgs}. Related to the expression (\ref{proto}), as far as the predictions are concerned,  are  the  logarithmic supersymmetry-inspired models $V=V_0(1+c (\ln \phi/\mu))$ obtained  for $p\rightarrow 0$, and  the brane inflation models obtained for $p=-4$. 
Potentials of the form $V(\phi)=V_0\left(1\pm c\phi^2\right)$ give a scale-independent spectral index.
In all these cases the predictions are \cite{lrreview} (for $p\neq 2$)

%\begin{eqnarray}
%n_s-1&=&-\left(\frac{p-1}{p-2}\right)\frac{2}{N},\nonumber\\
%\alpha_s&=&\left(\frac{p-1}{p-2}\right)\frac{2}{N^2}.
%\end{eqnarray}
\be
n_s-1=-\left(\frac{p-1}{p-2}\right)\frac{2}{N},\,\,\,\alpha_s=-\left(\frac{p-1}{p-2}\right)\frac{2}{N^2},
\ee
%where $N$ is the number of {\it e}-folds till the end of inflation. 
The running  can be written as
%
%\begin{eqnarray}
%\label{single}
%\alpha_s&=&-\frac{1}{2}\left(\frac{p-1}{p-2}\right)(n_s-1)^2\nonumber\\
%&=&-\frac{72}{25}\left(\frac{p-1}{p-2}\right)f_{\rm NL}^2\nonumber\\
%&\simeq& -6\left(\frac{p-2}{p-1}\right)\times 10^{-4},
%\end{eqnarray}

\begin{eqnarray}
\label{single}
\alpha_s&=&-\frac{1}{2}\left(\frac{p-2}{p-1}\right)(n_s-1)^2=
-\frac{72}{25}\left(\frac{p-2}{p-1}\right)f_{\rm NL}^2\nonumber\\
&\simeq& -6\left(\frac{p-2}{p-1}\right)\times 10^{-4},
\end{eqnarray}
where in the second passage we have made use of the known relation $f_{\rm NL}=-5(n_s-1)/12$ valid for single-field models \cite{maldacena}
and the numerical estimate has been done  using  the Planck central value for $(n_s-1)$ (as we will do in the following).
The factor $(p-2)/(p-1)$ is of order unity and  
the most favorable case is  achieved for the logarithmic models, for which  $p\simeq 0$ and $\alpha_s\simeq 1.2\times 10^{-3}$. 

Since the SPHEREx measurements of the power spectrum and the bispectrum may provide  $1\sigma$ level detection of the running spectral index
at the level of $6.5\times 10^{-4}$ \cite{dore},  a non-detection of the running  will  rule out  models with $p<-12$ at the $1\sigma$ level.
% and {\it all} models with
%$-\infty<p<1$. 
%The very same conclusion is reached for models of the kind $V(\phi)=V_0\left(1+c\phi^p\right)$ with $p$ an integer
%$\geq 3$ (tree-level self-coupling) or $\leq -1$ (dynamical symmetry breaking) as
%the same expression (\ref{single}) holds.   
On the other hand a  $2\sigma$ detection of the running would require
a running larger than $1.3\times 10^{-3}$, which is almost impossible to get  in single-field slow-roll models of inflation if no tensors are observed, unless $(n_s-1)$ is redder than the  Planck central value by more than two Planck standard deviations. Of course, a large running can be obtained by abandoning the slow-roll regime \cite{sloth}, but in this case no correlation with the non-Gaussianity is expected.

\vskip 0.2cm
\noindent
{\em Multi-field inflation.}--- In the single-field inflation scenario the total curvature perturbation is generated during inflation and remains constant on super-Hubble scales. However, in many inflationary  models  there are usually a plethora of scalar fields
 which may play a 	significant role during inflation.  Indeed, if these fields do not dominate the energy density during inflation, but are light enough to  be quantum-mechanically excited during the de Sitter stage, they   provide a source of isocurvature perturbation.  

If the curvature perturbation associated to the field driving inflation is suppressed, the total curvature perturbation may be originated
from such isocurvature perturbation. 
%All these mechanisms are operative if the contribution to the curvature perturbation coming from the degree of freedom driving inflation is suppressed. 
From Eq. (\ref{pscalar}) one deduces that this is naturally achieved
if the Hubble rate is small: no significant amount of tensor modes are expected in the case in which the seeds of the LSS and CMB anisotropies are due to a light field other than the inflaton (this is in fact not a no-go theorem \cite{zaf}, but it is a very well-educated guess).

We will concentrate on the most studied example of   such a mechanism, the so-called  
  curvaton 
mechanism
\cite{curvaton1,LW,curvaton3,LUW}, even though our findings  hold as well for 
other mechanisms  \cite{gamma1,gamma2,end1,end2}.
%the so-called modulated decay scenario \cite{gamma1,gamma2,gamma3} as well as in the case in which the  dominant contribution to the primordial curvature perturbation  is generated at the end of inflation \cite{end1,end2}.
%Another  mechanism is represented by the so-called  modulated decay scenario \cite{gamma1,gamma2,gamma3}, 
%in which the decay rate of the inflaton field has    super-Hubble spatial
%fluctuations due to the fact that the decay rate depends on a light scalar field. At the time of reheating after inflation,  
%this originates a curvature  perturbation
%in  the final reheating temperature
%in different regions of the universe. A third example  is the generation of  the dominant contribution to the primordial curvature perturbation  at the end of inflation which takes place at different times in different patches due to the presence of a light field \cite{end1,end2}.
%All these mechanisms are operative if the contribution to the curvature perturbation coming from the degree of freedom driving inflation is suppressed. 
%From    Eq. (\ref{pscalar}) one deduces that this is naturally achieved
%if the Hubble rate is small: no significant amount of tensor modes are expected in the case in which the seeds of the LSS and CMB anisotropies are due to a light field other than the inflaton.
In the curvaton mechanism, during inflation 
the curvaton energy density  is 
negligible and isocurvature perturbations with
a nearly flat spectrum are produced in the curvaton field $\sigma$.

%\be
%{\cal P}_{\delta\sigma}(k) = \left(\frac{H_*}{2\pi}\right)^2.
%\ee
After the end
of inflation, 
the curvaton field oscillates during some radiation-dominated era,
causing its energy density to grow and 
thereby converting the initial isocurvature into curvature 
perturbation. 
The comoving curvature perturbation $\zeta$ can be computed through the $\delta N$-formalism \cite{deltaN} 
$\zeta(\vec{x},t)=N(\vec{x},t)-N(t)$, where $N(\vec{x},t)$ is the amount of expansion along the worldline of a comoving observer from a spatially
flat slice at some initial time  to a generic slice at time $t$. 
Since 
 the number of {\it e}-folds from the end of inflation
to the beginning of the oscillations is unperturbed
because  the radiation energy density $\rho_{\rm rad}=(\rho_{\rm tot}-\rho_\sigma)$ dominates during
this time, for  the curvaton case one can
redefine $N$ as the number of {\it e}-folds from the beginning
of the sinusoidal oscillations (that is when the quadratic part of the curvaton potential with mass $m$ starts dominating)  
to the curvaton decay  \cite{lythrod}
\be
N\left[\rho(t_{\rm d}),\rho(t_{\rm osc}),\sigma_*\right]=\frac{1}{4}\ln\frac{\rho_{\rm rad}(t_{\rm osc})}{\rho_{\rm tot}(t_{\rm d})-\rho_\sigma(t_{\rm d})},
\ee
where

\be
\label{as}
\rho_\sigma(t_{\rm d})=\frac{1}{2}m^2g^2(\sigma_*)\left(\frac{\rho_{\rm tot}(t_{\rm d})-\rho_\sigma(t_{\rm d})}{\rho_{\rm rad}(t_{\rm osc})}\right)^{3/4}.
\ee
We have highlighted the role of the function 
$g=g(\sigma_*)$. It represents  
the initial amplitude of curvaton oscillations which is  some
function of the field value at the Hubble exit. The generic function $g(\sigma_*)$ parametrizes the  non-linear evolution on large scales if the
curvaton potential deviates from a purely quadratic potential with mass $m$
away from its minimum \cite{nurmi,kor}.

The resulting power spectrum of the scalar perturbations reads
\cite{lythrod}

\be
\label{a}
{\cal P}_s(k)=\frac{4r^2_{\rm d}}{9}\left(\frac{H_*}{2\pi}\right)^2\left(\frac{g'}{g}\right)^2,
\ee
where 

\be 
r_{\rm d}=\frac{3\rho_\sigma(t_{\rm d})}{3\rho_\sigma(t_{\rm d})+4\rho_{\rm rad}(t_{\rm d})}
\ee
 parametrizes  the curvaton energy density at the time of curvaton decay with respect to the radiation component.

%where
%-----FIGURE---------%
\begin{figure*}
\centering 
\resizebox{0.47\textwidth}{!}{\includegraphics{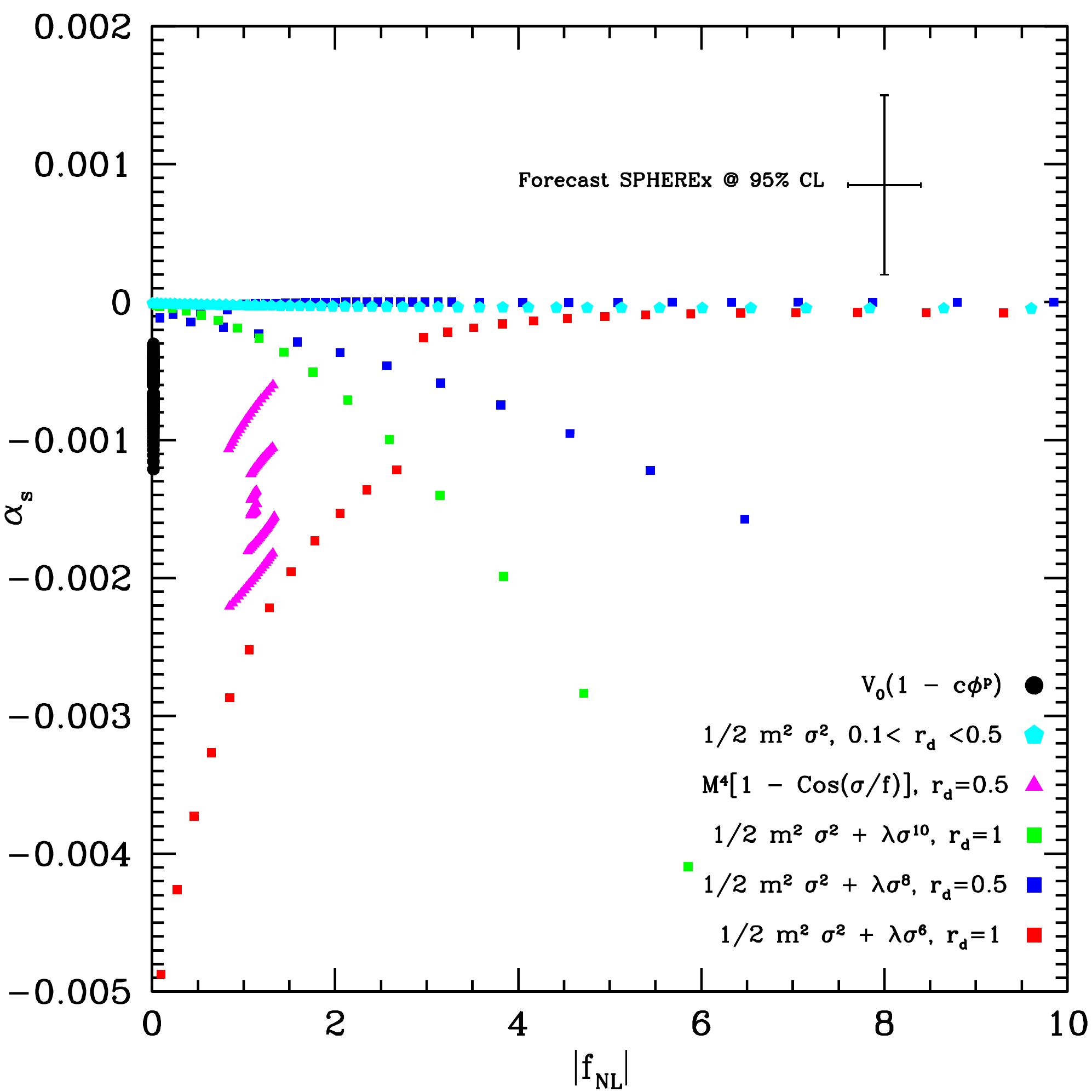}}
\resizebox{0.47\textwidth}{!}{\includegraphics{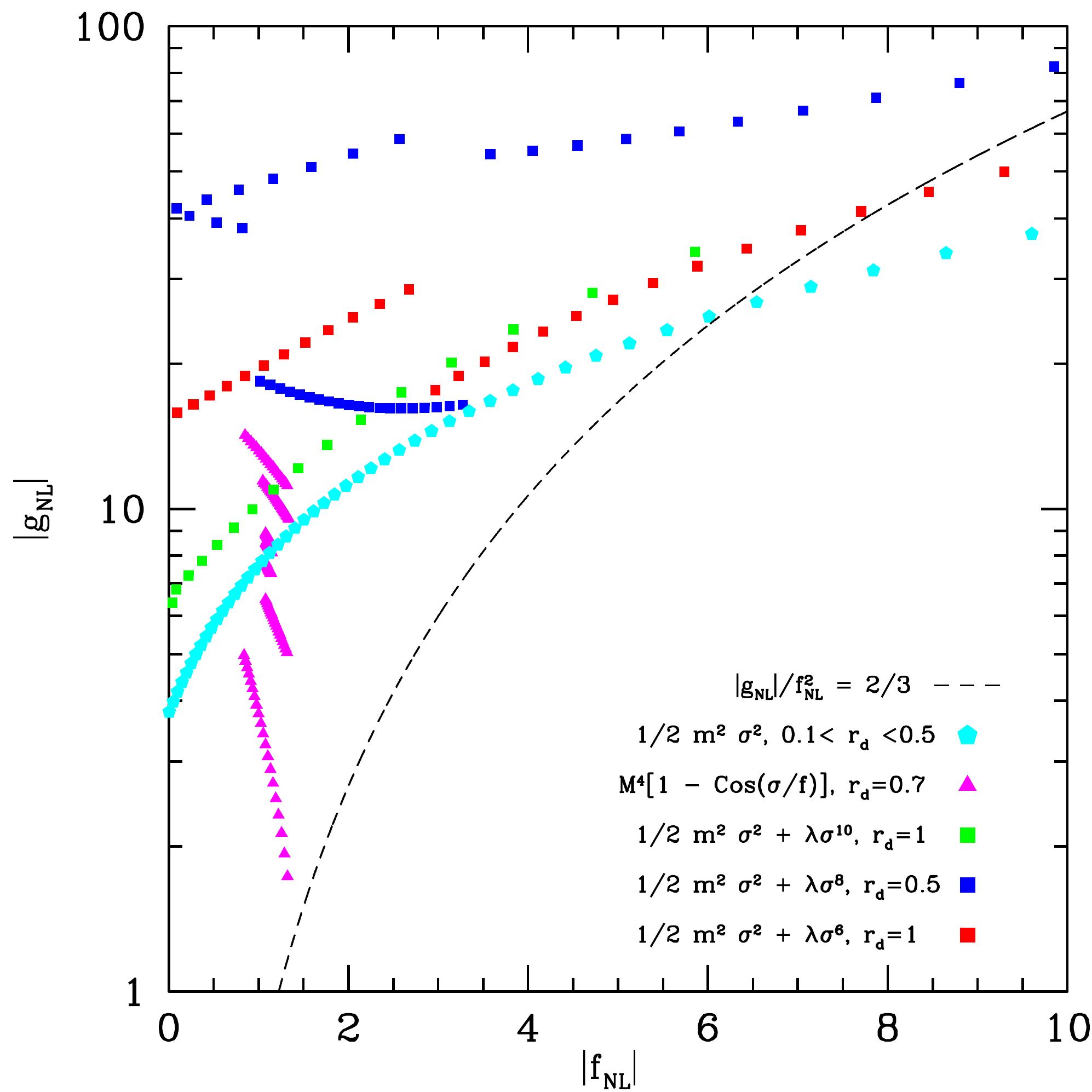}}
\caption{Left: The running  $\alpha_s$ as a function of $f_{\rm NL}$ for a range of curvaton models. Right: Non-linearity parameter $g_{\rm NL}$ as a function of $f_{\rm NL} $ for the same models as in the left plot.}
\label{fig:all}
\end{figure*}

%-----FIGURE---------%

%Taking derivatives with respect to $\sigma_*$ (indicated in the following by the  prime ${}^{'}$) at fixed $\rho(t_{\rm d})$ and  $ \rho_{\rm rad}(t_{\rm osc})\simeq \rho(t_{\rm osc})$, one finds 
% the power spectrum of the curvature perturbation reads \cite{lythrod}
%
%\be
%\label{a}
%{\cal P}_s(k)=\frac{4r^2_{\rm d}}{9}\left(\frac{H_*}{2\pi}\right)^2\left(\frac{g'}{g}\right)^2,
%\ee
%where $r_{\rm d}=3\rho_\sigma(t_{\rm d})/(3\rho_\sigma(t_{\rm d})+4\rho_{\rm rad}(t_{\rm d}))$ parametrizes  the total energy density at the time of curvaton decay.  
%
We neglect for the time being  
 the scale-dependence of $H_*$ 
% (one can easily show that the contribution to it from curvaton is suppressed with respect to the remaining terms and the one from  the inflaton can be taken arbitrarily small).   
 and  write  
 $\d /\d\ln k=H_*^{-1}\d/\d t_*=(\dot{\sigma}_*/H_*)\partial/\partial \sigma_*$. We then 
%\ee
%\be
%\frac{\d }{\d\ln k}=\frac{1}{H_*}\frac{\d}{\d t}=\frac{\dot{\sigma}_*}{H_*}\frac{\partial}{\partial \sigma_*},
%\ee
find

\be
(n_s-1)=2\frac{\dot{\sigma}_*}{H_*}\frac{g'}{g}\left(\frac{g'' g}{g'^2}-1\right)+2\frac{\dot{\sigma}_*}{H_*}\frac{r'_{\rm d}}{r_{\rm d}},
\ee
and
\begin{eqnarray}
\label{fund1}
\alpha_s&=&2\left(\frac{\dot{\sigma}_*}{H_*}\frac{g'}{g}\right)^2\left(\frac{g''' g^2}{g'^3}-\frac{g''^2g^2}{g'^4}-\frac{g''g}{g'^2}+1\right)\nonumber\\
&+&2\left(\frac{\dot{\sigma}_*}{H_*}\right)^2\left(\frac{r''_{\rm d}}{r_{\rm d}}-\frac{r'^2_{\rm d}}{r^2_{\rm d}}\right)\nonumber\\
&-&\frac{2}{3}\frac{V''}{H_*^2}\frac{\dot{\sigma}_*}{H_*}\left[\frac{g'}{g}\left(\frac{g'' g}{g'^2}-1\right)+\frac{r'_{\rm d}}{r_{\rm d}}\right],
\end{eqnarray}
%\begin{eqnarray}
%(n_s-1)&=&2\frac{\dot{\sigma}_*}{H_*}\frac{g'}{g}\left(\frac{g'' g}{g'^2}-1\right)+2\frac{\dot{\sigma}_*}{H_*}\frac{r'_{\rm d}}{r_{\rm d}},\nonumber\\
%\alpha_s&=&2\left(\frac{\dot{\sigma}_*}{H_*}\frac{g'}{g}\right)^2\left(\frac{g''' g^2}{g'^3}-\frac{g''^2g^2}{g'^4}-\frac{g''g}{g'^2}+1\right)\nonumber\\
%&+&2\left(\frac{\dot{\sigma}_*}{H_*}\right)^2\left(\frac{r''_{\rm d}}{r_{\rm d}}-\frac{r'^2_{\rm d}}{r^2_{\rm d}}\right)\nonumber\\
%&-&\frac{2}{3}\frac{V''}{H_*^2}\frac{\dot{\sigma}_*}{H_*}\left[\frac{g'}{g}\left(\frac{g'' g}{g'^2}-1\right)+\frac{r'_{\rm d}}{r_{\rm d}}\right],\nonumber\\
%&&
%\end{eqnarray}
with $r'_{\rm d}=(2g'/3g)r_{\rm d}(1-r_{\rm d})(3+r_{\rm d})$.
%\be
%r'_{\rm d}=\frac{2}{3}\frac{g'}{g}r_{\rm d}(1-r_{\rm d})(3+r_{\rm d}).
%\ee
The expressions for the  quadratic and cubic non-Gaussianities are \cite{NGreview,sasakiwands}

\begin{eqnarray}
f_{\rm NL}&=&\frac{5}{4r_{\rm d}}\left(1+\frac{g'' g}{g'^2}\right)-\frac{5}{3}-\frac{5r_{\rm d}}{6},\nonumber\\
g_{\rm NL}&=&\frac{25}{54}\left[\frac{9}{4r_{\rm d}^2}\left(\frac{g'''g^2}{g'^3}+3\frac{g'' g}{g'^2}\right)
-\frac{9}{r_{\rm d}}\left(1+\frac{g'' g}{g'^2}\right)
\right.\nonumber\\
%&-&\frac{9}{r_{\rm d}}\left(1+\frac{g'' g}{g'^2}\right)\nonumber\\
&+&\left.\frac{1}{2}\left(1-9\frac{g'' g}{g'^2}\right)+10r_{\rm d}+3r^2_{\rm d}\right].
\end{eqnarray}
Notice that $r_{\rm d}$ depends on $\sigma_*$ through $\rho_\sigma(t_{\rm d})$ in Eq. (\ref{as}).  This dependence is essential to obtain the correct  $f_{\rm NL}$ and $g_{\rm NL}$ given above \cite{lythrod}.
%By re-expressing  the spectral index as 
%
%\be
%(n_s-1)=\frac{8}{5}r_{\rm d}\frac{\dot{\sigma}_*}{H_*}\frac{g'}{g}f_{\rm NL},
%\ee

We can recast the 
running of the spectral index in rather appealing and compact form
\be
\label{fund}
\alpha_s=\frac{3}{4}(n_s-1)^2\left(\frac{g_{\rm NL}}{f_{\rm NL}^2}-\frac{2}{3}\right)-\frac{V''}{3 H_*^2}(n_s-1).
\ee
The same expression can be  derived more directly in terms of derivatives of the number of {\it e}-folds using the $\delta N$-formalism \cite{deltaN}, but we have preferred to adopt  a slightly more involved derivation to stress the importance of the  non-linearities in the curvaton potential.

Eq. (\ref{fund}) is the main result of this paper and indicates that the running of the spectral index can be sizable if
 $g_{\rm NL}\gsim f_{\rm NL}^2$.  In general, this is true  in those 
   models for which $r_{\rm d}\simeq 1$  and $g''g/g'^2\simeq 1$ (or $r_{\rm d}\lsim 1$  and $g''g/g'^2\simeq -1$) when $f_{\rm NL}$ becomes small. This can happen  if the curvaton potential has non-linearities in it  \cite{nurmi1,by}.
As a rule of thumb, being $3(n_s-1)^2/4\simeq 9\times 10^{-4}$, we deduce that SPHEREx could detect a running of the spectral index 
 at   $2\sigma$ if $g_{\rm NL}/ f_{\rm NL}^2={\cal O}(\pm 2)$. Since SPHEREx can detect at  $1\sigma$-level a local   $f_{\rm NL}$ larger than 0.2, the situation seems optimistic.

Consider for instance the case in which   the curvaton potential has a self-interacting piece
 $\lambda\sigma^4$, where  $\lambda$ is a dimensionless coupling,  besides the quadratic term. 
%One can show that $g(\sigma_*)\sim \sigma_*^{3/4}$  if the non-linear part of the potential dominates during inflation \cite{nurmi1,by}. For $r_{\rm d}\lsim 1$, one obtains 
%% 
% \be
%n_s-1=-\frac{4}{3}\lambda\frac{\sigma_*^{2}}{H_*^2},\,\,\,\,\,\,\,
%f_{\rm NL}=\frac{5}{6r_{\rm d}},
%\ee
%and
%\be
%g_{\rm NL}=-\frac{2}{3}f^2_{\rm NL},\,\,\,\,
%\alpha_s=2(n_s-1)^2\simeq 3.6\times 10^{-3}.
%\ee
 %\begin{eqnarray}
%n_s-1&=&-\frac{4}{3}\lambda\frac{\sigma_*^{2}}{H_*^2},\nonumber\\
%f_{\rm NL}&=&\frac{5}{6r_{\rm d}},\nonumber\\
%g_{\rm NL}&=&-\frac{2}{3}f^2_{\rm NL},\nonumber\\
%\alpha_s&=&2(n_s-1)^2\simeq 3.6\times 10^{-3}.
%\end{eqnarray}
% Next we  consider a self-interacting non-renormalizable potential for the curvaton of the form $V(\sigma)=\lambda_n\sigma^n$, with $n>5$.
  In the limit in which the non-linear term dominates over the quadratic piece during inflation 
$g(\sigma_*)\sim \sigma_*^{3/4}$ \cite{nurmi1,by}. One obtains

\be
n_s-1=\frac{4}{3}\lambda(r_{\rm d}(2+r_{\rm d})-1)\frac{\sigma_*^2}{H_*^2}, \,\, 
f_{\rm NL}=\frac{5}{6r_{\rm d}}\left(1-r^2_{\rm d}-2r_{\rm d}\right).
\ee
%and
%\be
%f_{\rm NL}=\frac{5}{2r_{\rm d}}\frac{(6-n)}{(10-n)},\,\,\,\,\,\, g_{\rm NL}=\frac{2}{3}\frac{(2-n)}{(6-n)}f^2_{\rm NL}
%\ee
%and
%
%\be
%\alpha_s=-2\frac{(n-2)}{(n-6)}(n_s-1)^2.
%\ee
%%\begin{eqnarray}
%%n_s-1&=&\lambda_n\,n(n-6)\frac{\sigma_*^{n-2}}{6H_*^2},\nonumber\\
%%f_{\rm NL}&=&\frac{5}{2r_{\rm d}}\frac{(6-n)}{(10-n)},\nonumber\\
%%g_{\rm NL}&=&\frac{2}{3}\frac{(2-n)}{(6-n)}f^2_{\rm NL},\nonumber\\
%%\alpha_s&=&-2\frac{(n-2)}{(n-6)}(n_s-1)^2.
%%\end{eqnarray}
For instance, for  $r_{\rm d}=1/3$ we get  a red tilt, 
%a red tilt   $(n_s-1)=-(5\lambda/6)(\sigma_*^3/H_*^2)$, 
$f_{\rm NL}=5/9$, 
$g_{\rm NL}\simeq -32 f_{\rm NL}^2$,   and  $\alpha_s\simeq -13\times 10^{-3}$.  This running can certainly be measured by    EUCLID and SPHEREx.

To describe further  the correlation between a sizable running $\alpha_s$ and the level of non-Gaussianity, we present some illustrative examples obtained numerically in Fig. \ref{fig:all}.  For completeness, we have restored the scale-dependence of $H_*$ 
through an  inflaton potential  of the form $V(\phi)\sim {\rm exp}(\mu\phi/\mp^2)$, which  provides a  constant contribution $(-\mu^2/\mp^2)$ to  $(n_s-1)$.
%One should use in such a case the expression (\ref{fund1}) for the running.
The various points have been obtained finding the function $g(\sigma_*)$,  scanning over the various parameters and using the general expression (\ref{fund1}). We have systematically checked that all the observables, such as   the amplitude of the power spectrum, are in agreement with the current observations and that the contribution to the running from the $V''$-term in Eq. (\ref{fund}) and from the scale-dependence of $H_*$  is smaller than the one 
coming from the ratio $g_{\rm NL}/f^2_{\rm NL}$. 
The $2\sigma$ SPHEREx forecasted errors
for $\alpha_s$ and $f_{\rm NL}$ are also shown. On the right panel we present  the corresponding values of $g_{\rm NL}$ as a function of  $f_{\rm NL}$. The dash line illustrates that the running is mainly due to the correlation with the  non-Gaussianity.

Finally, let us consider  the case of the modulated reheating scenario.  The curvature perturbation is generated during reheating  if the inflaton  decay rate $\Gamma=\Gamma(\sigma)$ is a function of a light field $\sigma$. The corresponding power spectrum and non-Gaussianities are \cite{sasakiwands}
 
 \begin{eqnarray}
{\cal P}_s(k)&=&\left(\frac{H_*}{18\pi}\right)^2\left(\frac{\Gamma'}{\Gamma}\right)^2,\,\,
f_{\rm NL}=5\left(\frac{\Gamma'' \Gamma}{\Gamma'^2}-1\right),\nonumber\\
g_{\rm NL}&=&\frac{100}{3}\left(1-\frac{3}{2}\frac{\Gamma'' \Gamma}{\Gamma'^2}+\frac{1}{2}\frac{\Gamma'''\Gamma^2}{\Gamma'^3}\right).
\end{eqnarray}
One can easily show that the analytical expression  for the running is  exactly the same as in Eq. (\ref{fund}). Now, at the time of reheating the 
field $\sigma$ will have a value $\sigma_{\rm reh}=\sigma_{\rm reh}(\sigma_*)$ and the 
plausible dependence of the  decay rate is  $\Gamma(\sigma_{\rm reh})\propto \sigma_{\rm reh}^q$ ($q>0$).
If the potential for the light field is quadratic, then $\sigma_{\rm reh}=\sigma_*$ and we find
$g_{\rm NL}/ f_{\rm NL}^2=4/3$: the first term in the running turns out to be of the order of $6\times 10^{-4}$, too small to be detected. 
Nevertheless, if there is a  non-linear relation between $\sigma_{\rm reh}$ and $\sigma_*$   due to the fact that  the
potential of  the field  $\sigma$ deviates from a purely quadratic potential, then one can easily increase the ratio $g_{\rm NL}/ f_{\rm NL}^2$ and make the running detectable.

% \be
%\alpha_s=6(n_s-1)^2\simeq 7\times 10^{-3}.
%\ee
%
 
%In the case of the modulated reheating scenario the curvature perturbation is generated when during reheating  if the inflaton  decay rate $\Gamma=\Gamma(\sigma)$ is a function of a light field $\sigma$. The corresponding power spectrum and non-Gaussianities are \cite{sasakiwands}
% 
% \begin{eqnarray}
%{\cal P}_s(k)&=&\frac{1}{36}\left(\frac{H_*}{2\pi}\right)^2\left(\frac{\Gamma'}{\Gamma}\right)^2,\nonumber\\
%f_{\rm NL}&=&-5\left(1-\frac{\Gamma'' \Gamma}{\Gamma'^2}\right),\nonumber\\
%g_{\rm NL}&=&\frac{100}{3}\left(1-\frac{3}{2}\frac{\Gamma'' \Gamma}{\Gamma'^2}+\frac{1}{2}\frac{\Gamma'''\Gamma^2}{\Gamma'^3}\right).
%\end{eqnarray}
%The analytical expression  for the running is  exactly the same as in Eq. (\ref{fund}). Now, at the time of reheating the 
%field $\sigma$ will have a value $\sigma_{\rm reh}=\sigma_{\rm reh}(\sigma_*)$ and the 
%plausible dependence of the  decay rate is  $\Gamma(\sigma_{\rm reh})\propto \sigma_{\rm reh}^n$.
%If the potential for the light field is quadratic, then $\sigma_{\rm reh}=\sigma_*$ and we find
%$g_{\rm NL}/ f_{\rm NL}^2=4/3$: the first term in the running turns out to be of the order of $6\times 10^{-4}$, too small to be detected. 
%Nevertheless, if there is a  non-linear relation between $\sigma_{\rm reh}$ and $\sigma_*$   due to the fact that  the
%potential of  the field  $\sigma$ deviates from a purely quadratic potential, then one can easily increase the ratio $g_{\rm NL}/ f_{\rm NL}^2$ and make the running detectable.
%
\vskip 0.2cm
\noindent
{\em Conclusions.}---In this paper we have stressed that a sizable running  of the spectral index   is typically  connected  
to the non-Gaussianity in the case in which the curvature perturbation is due to the presence of more than one scalar field during inflation. A future  detection of a significant running  of the spectral index  and non-Gaussianity may tell us a lot about the 
true mechanism giving rise to the  primordial perturbations. Dedicated searches aiming at measuring with great accuracy the running 
and  non-Gaussianity are of vital importance.

\vskip 0.2cm
\noindent
{\em Acknowledgements.}--- When completing this work, the paper \cite{new} appeared making also the point that the   running of the scalar spectral index is correlated with the statistical properties of non-Gaussianities. We thank the authors of Ref. \cite{new} for correspondence. We also thank O. Dor\' e for discussions on the capabilities of SPHEREx and M. Sloth for comments on the draft. 
 M.B. ackowledges support by the Swiss National Science Foundation. A.K. thanks the Aristeia II Action of the Operational Programme Education and
Lifelong Learning,  the European Social Fund (ESF), and the European Union’s Seventh Framework Programme (FP7/2007-2013) under REA grant agreement no.
329083. A.R. is supported by the Swiss National Science Foundation (SNSF), project ‘The non-Gaussian Universe”
(project number: 200021140236).

\end{document}